\documentstyle[aps,prl]{revtex}
\begin{document}
\draft
\twocolumn[\hsize\textwidth\columnwidth\hsize\csname@twocolumnfalse\endcsname 
\title{Spin-Peierls transition in  NaV$_2$O$_5$ in high magnetic
fields}
 
\author{S. G. Bompadre, Arthur F. Hebard, and Valeri N. Kotov}

\address{Department of Physics, Box 118440, 
University of Florida, Gainesville, FL 32611-8440, USA}

\author{Donavan Hall}

\address{National High Magnetic Field Laboratory, Tallahassee, FL
  32310, USA}

\author{G. Maris, J. Baas and T.T.M. Palstra}

\address{Solid State Chemistry Laboratory, Materials Science Centre, 
University of Groningen, 9747 AG Groningen, The Netherlands}

\maketitle
 
\begin{abstract}
We  investigate the magnetic field dependence of the spin-Peierls
 transition in NaV$_2$O$_5$ in the field range
 16T-30T. The transition temperature
 exhibits a very weak variation with the field, suggesting a novel
 mechanism for the formation of the  spin-Peierls state.
 We argue that a charge ordering transition accompanied by singlet
 formation is consistent with our observations.  
\end{abstract}

]

 The Peierls instability takes place in one-dimensional
 systems and can give rise to complex and fascinating behavior. In 
itinerant electronic systems the instability is driven by the 
coupling of electrons to the phonons of the lattice
\cite{peierls}. Any coupling 
at $T=0$ leads to the formation of the Peierls state which is characterized by 
charge ordering (gap in the electronic spectrum) and a finite lattice 
distortion.
Similar phenomena occur in  purely insulating spin
 systems,  where the spin-phonon coupling is responsible 
for the formation
 of a singlet ground state with neighboring spins  pairwise bound into
 singlets \cite{bray}. The spin-Peierls ground state shows a
 characteristic gap in the excitation spectrum and has been observed
 in a variety of organic compounds,
 such as (TTF)[CuS$_{4}$C$_{4}$(CF$_{3}$)$_{4}$] \cite{bray}.
 At high temperatures these materials behave as non-interacting
 Heisenberg chains, while below the transition temperature, T$_{c}$,
 the magnetic exchange acquires an alternating component.
 In 1993 the first inorganic
 spin-Peierls compound CuGeO$_{3}$ was discovered \cite{cugeo3}
 with  T$_{c}$ $\approx 14 K$. This material, like its organic
predecessors, shows a characteristic 1d Heisenberg
(Bonner-Fisher)-like magnetization at high T  
 with a sharp drop at T$_{c}$, indicating a non-magnetic
 ground state.  
 Very  recently, a second inorganic compound, NaV$_2$O$_5$  
was shown to behave as a spin-Peierls material with T$_{c}$ $\approx 34$K
\cite{first}. The properties of NaV$_2$O$_5$ however have proven
 to be  quite controversial, thus stimulating
 the research reported in this letter. 
                  
 Magnetic susceptibility measurements of NaV$_2$O$_5$
 indicate a transition
 to a non-magnetic phase  at T$_{c}$ \cite{first,weiden}.
 This can be understood within the framework of a 
 spin-phonon coupling driven transition on a Heisenberg chain
 \cite{bray}. The antiferromagnetic exchange, J, was estimated to be
 J$\approx$560K.  The low-temperature structure which is assumed in
 this interpretation of the data, is that of magnetic chains
 formed by the spin 1/2 V$^{4+}$ ions along the crystalline b-axis, 
  separated by spinless V$^{5+}$ chains. This scenario implies  a
  lattice distortion in one direction only.
 However, recent experiments have shown that the above picture  
 is not satisfactory. 
 X-ray diffraction measurements indicated that the system should be viewed
 as a quarter-filled ladder made of  V$^{4.5+}$ chains \cite{smolinski,meetsma},
 meaning that a spin of 1/2 is not attached to a single V ion, but rather
 to a rung of the ladder, i.e. a V-O-V orbital. 
 Subsequent NMR \cite{ohama} analysis revealed that below T$_{c}$,
 two inequivalent types of V sites - V$^{4+}$ and V$^{5+}$ appear,
 suggesting that charge ordering occurs in the spin-Peierls phase.
 Charge disproportionation  leaves room for period doubling 
in more that one crystallographic direction, consistent
 with additional X-ray \cite{ravy,palstra} and NMR \cite{fagot} studies.
 These works suggest that lattice distortion takes place in the (a,b) plane
 (where b is the direction along the chains and a is perpendicular to the 
chains). 
 A number of theoretical  studies \cite{mostovoy,seo,horsch,thalmeier,riera,mazumdar}
 have addressed the possibility of charge ordering in 1/4 filled
 systems, where both electron-lattice and
 electron-electron interactions are included. 
The most probable scenario at present seems to be the
 ``zig-zag'' order proposed in Ref. \cite{mostovoy} where
 the charge density (i.e. the sites
V$^{4.5\pm\delta}$ with deviation $\delta$ from the average
 valence) is distributed in a zig-zag fashion along
 the ladder direction. As emphasized in Ref.\cite{mostovoy} 
the Coulomb repulsion in combination with the electron-lattice interaction 
 can drive such a transition, while the formation
 of a spin singlet ground state ``follows'' the charge order.  
 Charge modulation is consistent 
with the analysis of the observed magnetic excitation spectra 
\cite{groslemmens},
 Raman spectra \cite{raman}, as well as the anomalies in the thermal 
conductivity
\cite{vasiliev} and the dielectric constant \cite{smirnov} at T$_{c}$.
 
 The present work attempts to gain further insight into the nature of
 the spin-Peierls transition in NaV$_2$O$_5$ by addressing the 
magnetic field dependence
 of the transition temperature in very high fields. Previous studies in
 fields up to 5.5T \cite{weiden} have found behavior  consistent
 with the theoretical predictions  and  similar to 
  the  previously known spin-Peierls compounds \cite{bray}. 
 However, subsequent measurements in higher fields, up to 14T \cite{fertey}
and 16T \cite{schnelle}, have found much weaker field dependence. These
 experiments were based on a determination of T$_{c}$ from
 the changes in the elastic constants \cite{fertey} and the specific
 heat \cite{schnelle},
 unlike the measurement in Ref. \cite{weiden} which determined T$_{c}$
 from the drop in the magnetization. 


 In this work we have measured the magnetization of two NaV$_2$O$_5$
 single crystals in magnetic fields from 16T to 30 T.  The crystals were grown 
by high temperature solution growth from a vanadate mixture flux.  The
masses of the samples investigated were 1.9 and 3.1 mg respectively
and they had irregular parallelepiped shapes with smooth, faceted faces.
The single crystals were characterized with an Enraf-Nonius CAD4 single 
 crystal diffractometer using Mo-radiation. The results of the structure refinement
were the same as reported earlier in Ref.\cite{meetsma}.
Magnetization was measured using a standard metal foil
cantilever beam magnetometer. The ``T'' shaped flexible cantilever beam was made 
from a 7.62 $\mu$m thick heat treated MP35N alloy.  
The dimensions of the ``T'' were
 approximately 8 mm on a side. The gap between the ``T''  and the parallel 
fixed reference electrode was approximately 800 $\mu$m.
The sample was mounted using a small amount of vacuum grease.  In the
 presence of a DC magnetic field the interaction of
 the magnetic moment of the sample with the field results in
 a force and/or torque, deflecting the beam and changing 
the capacitance between electrodes. A capacitance bridge was used
to monitor 
the changes in force (magnetization) for temperature sweeps in fixed field.  
Since MP35N is magnetic (typically 13.5 $\mu$emu/g at 78 K), the same
 bare cantilever was measured under the identical conditions (sweep
 direction and sweep rate) as the
 cantilever+sample combination to provide a background reference.  
The temperature dependence of the
 cantilever capacitance was compensated for in the same way. A room 
temperature measurement of the cantilever's sensitivity showed that a force
 of 3 nN could be resolved.

Cantilever displacement can arise from either a torque or a force on a
sample with a magnetic moment. When the sample is at field center,
where the field 
gradient is zero, then torque ($ \propto{\bf m} \times {\bf B}$) will
dominate. Strictly speaking, if the sample is isotropic and there are
no shape factors, then there is no
 torque on the cantilever for fields applied along the direction of
 displacement (perpendicular to the sample). On the other hand, 
when the sample is raised (or lowered) away from field center, the force term 
($F= m \;  dB/dz  \propto \; \chi \; B \; dB/dz$) will usually 
dominate, although torques can still be present. 
Figure \ref{fig.1} shows temperature sweeps taken for the 1.9 mg
sample at the three indicated fields in the
 legend. The cantilever was located in a position where the field
 gradient was maximum. The maximum field at this position (24T) is
 80$\%$ of the field center maximum (30T). 
The change in capacitance, $\Delta C$, which is proportional to the
change in magnetization, is calculated at each field by subtracting
the background trace (cantilever alone) from the sample trace
(sample+cantilever).
This quantity is divided by B$^{2}$ and plotted as the ordinate in Fig. 
\ref{fig.1}. As seen from the figure, the data scale reasonably well
for the three different fields, confirming the B$^{2}$ dependence
expected from both torque and force contributions. To accentuate the
small shift in the transition temperature, we plot in the inset the
derivative of $\Delta C/B^{2}$ with respect to the temperature. From
the position of the peaks we can determine the field-dependent
transition temperature. 

To measure $T_{c}$ at the maximum field of 30T, the sample was
placed at field center (sensitive to the torque only) and the data
collected and analyzed as described above. A similar scaling with
B$^{2}$ was observed. Figure \ref{fig.2} shows the derivative of
$\Delta C$ with respect to
temperature at field center.  Plotted in this way, the shift in
$T_{c}$ can be clearly seen. Data similar to those plotted in
Figs. 1 and 2 were obtained for a second sample with mass 3.1 mg, 
and for reversed fields.  In all cases the shifts in
$T_{c}$ were equal to or less than the shifts shown in Fig. \ref{fig.2}. 
In  Fig. \ref{fig.2}  (inset) we also show the results of magnetization measurements 
at low fields using a
commercial SQUID magnetometer (MPMS7). The singlet formation at $T_{c}$ is clearly observable,
 but no shift of $T_{c}$ can be observed within measurement accuracy
 in fields up to 5T, in agreement with previous work  \cite{first}.

In Fig. \ref{fig.3} we present our high field data for the variation of T$_{c}$
 in terms of $\Delta T_{c}/T_{c}(0) = T_{c}(H)/T_{c}(0) -1$
 and the square of the scaled magnetic field $h = g \mu_{B}H/2k T_{c}(0)$
\cite{gfactor}.
This scaling is expected in  spin-Peierls systems, and for
 small fields $h \ll 1$, the relative variation of $T_{c}$ should be quadratic
\cite{bray}:

\begin{equation}
\Delta T_{c}/T_{c}(0) = - \alpha h^{2}.
\end{equation}
The data of Fig. \ref{fig.2} follows this  dependence quite well, and
 we estimate $\alpha_{exp} \approx 0.072(8)$. The value $T_{c}(0)$ was
 not measured directly but was estimated from an extrapolation to
 zero field of the quadratic
 dependence of $T_{c}$ vs. $H$ to be  $T_{c}(0)=34.2K$.
This value is close to published values and to
 the $T_{c}(0)$ measured by us using SQUID magnetometer measurements
 of the magnetization of a 40mg polycrystalline sample.
 The combination of our high field data and the lower field data of
 previous measurements 
 gives the variation of $T_{c}$ over a large range of magnetic
 field and  shows a very weak dependence. 
 In contrast, the ``conventional'' inorganic spin-Peierls compound
CuGeO$_{3}$ exhibits a much stronger field dependence with
 $\alpha = 0.39$ \cite{hasepoirier}, in good agreement with
 the theory.
 The theoretical values of $\alpha_{SP}$ predicted for
 the spin-Peierls transition are $\alpha_{SP}=0.44$ or $0.36$,
 depending on the way interaction effects are taken into account
\cite{bray}.
 The first, larger number corresponds to the Hartree
 approximation for the interactions between the Jordan-Wigner
 fermions, representing the localized spins \cite{bulaevskii}. 
 The value $0.36$ is obtained by exact treatment of
 the correlation effects \cite{cross}, which is possible
 in the Luttinger liquid framework in one dimension \cite{luther}. 
 In both cases the characteristic scaling $H/T_{c}(0)$ which appears
 in Eq.(1) is due to the commensurate nature of the dimerized phase.
For large fields, corresponding to a 
 reduction of  $T_{c}$ by a factor of
$T_{c}/T_{c}(0)=0.77$, 
 a transition into an incommensurate
 phase is expected  to take place \cite{cross}.
 Such a transition is   less sensitive to  magnetic field  and has been
 observed in a variety of spin-Peierls materials \cite{bray,hasepoirier}.
 In NaV$_{2}$O$_{5}$ however, a transition into such a modulated phase
 does not seem to take place, since even in the highest field (30T), 
  $T_{c}(30T)/T_{c}(0)= 0.97$, which is very far from the expected
 incommensurate boundary.
 Notice that even in a field 
  as high as $30T$ the scaled ratio $h=0.59$ is quite small due to
 the large $T_{c}(0)$. 
 
We now discuss the possible sources for the difference
 between the measured value $\alpha_{exp}$ and the theoretically
 predicted one $\alpha_{SP}=0.36 \approx 5 \alpha_{exp}$ for spin-Peierls 
systems. 
In addition to this discrepancy, any theory  of NaV$_{2}$O$_{5}$
should also be able to explain
 the large value of the ratio $2\Delta/T_{c}(0) \approx 6$ ($\Delta
 \approx 100K$ \cite{ohama} being the spin gap), where a  
mean-field value of 3.52 might be expected.
       
As discussed in the introduction, a transition into a charge ordered
 state in a 1/4-filled system 
 is consistent with a number of recent experiments.
 Although it is not clear whether the charge density wave (CDW) 
precedes or forms simultaneously with the magnetically dimerized spin-Peierls
 state, it seems certain that the physics of charge ordering must be taken into
account.
Recent numerical work
 has shown \cite{riera,mazumdar} that CDW and spin-Peierls order
 can co-exist in quasi one-dimensional 1/4-filled electronic systems.
  If we assume that the 
 CDW formation is the driving force behind
 the opening of a spin gap, as argued in Ref.\cite{mostovoy},  
then the  ``charge'' part of the transition  will
 be mainly responsible for the $T_{c}(H)$ dependence.
 In a system of non-interacting electrons, undergoing a Peierls transition
  into a
(commensurate) CDW state, the decrease of  $T_{c}$ for small magnetic field
 (coupled to the electron spin via a Zeeman term) is also described
 by Eq.(1), but with $\alpha_{CDW}=0.21$ \cite{dieterich}.
 
Two effects, orbital coupling and electron-electron 
interactions, could further modify this result. 
Orbital effects are known to be present 
 when nesting is imperfect, and generally compete with the
 Pauli terms, producing a flatter dependence of $T_{c}$ on $H$,
 i.e. a further reduction of $\alpha_{CDW}$  \cite{gille}.
 However spin-orbit interactions lead to anisotropic variation of
 $T_{c}$ with respect to the magnetic field direction.
 In NaV$_2$O$_5$ this variation has been found to be extremely
 weak \cite{first,weiden,fertey,schnelle}, which is also confirmed
 in this work, and consequently the orbital effects can be ruled out
 as a source of the weak $T_{c}(H)$ dependence.
On the other hand, electron-electron interaction effects do not
reflect anisotropies and are  important in the formation and
 stabilization of a CDW state \cite{mostovoy,riera}.
 In general, the stability of the CDW depends on the strength of
 the electron-phonon coupling (which drives the transition) and on the on-site
 and nearest-neighbor Coulomb correlations \cite{riera}.
 
To demonstrate this latter point concerning strong correlation
 effects, we consider the simplified model
 of a Hubbard chain with an on-site
 repulsion $U$. We treat the phonons adiabatically,
 as in Ref.\cite{dieterich}, but take into account 
 the electron-electron interaction following Ref.\cite{luther},
 i.e. calculate the polarization bubble exactly for the Luttinger
 liquid. In this case it is known that $T_{c}(H=0)$ increases
 with respect to its value at $U=0$ \cite{chui}.
 For finite magnetic field we find,
 at $U \sim 2t$ (where $t$ is the bandwidth),  that
 the coefficient $\alpha$ drops to $\alpha_{CDW} \approx 0.15$,
 i.e. below the non-interaction value of 0.21. This is not
surprising and in fact is quite similar to the difference between
the mean-field and the exact treatment in the spin-Peierls case
($\alpha_{SP}=0.44, \, 0.36$, respectively). 
 The essence of the effect is in the different type of divergence in
 the polarization bubble with and without interactions.
 While in the free  case the polarization diverges
 logarithmically at small frequencies, in a Luttinger liquid
 the stronger, power law dependence sets in \cite{luther}, and
the Peierls instability is effectively enhanced.  
 Thus the interaction effects, being naturally more important
 for the CDW formation (compared to the spin-Peierls case),
 can produce a weaker $T_{c}(H)$ dependence. A more realistic
 calculation based on a Hamiltonian appropriate for
 NaV$_2$O$_5$ would be very desirable.

The orbital and interaction effects discussed above are, strictly
speaking, valid only
 for an isolated chain. It was  assumed that inter-chain 
 interactions are sufficiently strong to suppress the fluctuation
 effects, typically important in one-dimensional systems \cite{heinz}. 
 The fluctuations are known to reduce $T_{c}(0)$ below the mean-field
 value and cause a specific heat jump at the transition $\Delta c_{P}$
 several times the mean-field one. The large observed ratio
$\Delta/T_{c}(0)$ (twice the mean-field),
 in combination with a $\Delta c_{P}$ about ten times
 the mean-field value \cite{powell} suggest that fluctuations indeed
 could be important in NaV$_2$O$_5$. 
 At the same time one should have in mind that, due to the specific structure
 of NaV$_2$O$_5$, transverse interchain interactions are expected to play a
 crucial role in the stabilization of the ordered phase, in particular
 the formation of the spin gap and doubling of the period in the (a,b) plane
 \cite{ravy,fagot}. The vanadium displacements are nearly absent along
 the ladder direction (b-axis), and  largest perpendicular to the
 ladder direction both along the a- and c-axis \cite{palstra}.  
Thus it is not clear whether fluctuation effects have
to be necessarily invoked to explain the large $\Delta/T_{c}(0)$ ratio 
 in this material as is traditionally done, or whether the large 
$\Delta/T_{c}(0)$ ratio is intimately related to the anomalously weak 
variation of $T_{c}$ with field reported in this work.

We are grateful to A. Dorsey, S. Hershfield,  D. Maslov, C. Biagini
and S. Arnason for many stimulating discussions, and M. Meisel for
 critical reading of the manuscript.
 We are also appreciative of experimental advice given
by V. Shvarts and by the Users Support Group at the National High
Magnetic Field Laboratory (NHMFL). G.M. acknowledges the support
of the Netherlands Foundation for Fundamental Research on Matter with
financial aid from NWO. 
  A.F.H. and S.G.B. were supported
by the NSF funded In-House Research Program of the NHMFL in
Tallahassee and V.N.K. was supported by NSF Grant DMR9357474.

\vspace{2cm}

\begin{figure}
\caption
{Change in capacitance measured with a cantilever beam magnetometer
  off field center for a NaV$_2$O$_5$ single crystal in B = 16T, 20T,
  and 24T. $\Delta C$
  is proportional to the magnetization of the sample and has been
  normalized by the square of the magnetic field. The inset shows
  the derivative of the scaled, background subtracted data with
  respect to temperature.  $T_{c}$ is determined from 
  the position of the peaks.} 
\label{fig.1}
\end{figure}

\begin{figure}
\caption
        {Derivative of the unscaled capacitance readings (proportional
          to magnetization) with respect to temperature for the sample
          shown in Fig. 1 located at field center. 
          Inset shows low field SQUID magnetization measurements.}
\label{fig.2}
\end{figure}

\begin{figure}
\caption {
 Relative variation of T$_{c}$
 as a function of the scaled magnetic field
$h=g \mu_{B}H/2k T_{c}(0)$ (see text).
 The circles are our data
(numbers represent the values of the
 field  in Tesla), and the squares are data from Ref.
\protect \cite{schnelle},
 based on measurements of the specific heat jump at the transition.
}
\label{fig.3}
\end{figure}


\begin{references}
\bibitem{gfactor} The g-factor varies very weakly in 
 the three directions \cite{weiden,schnelle}, and we take $g=2$ in Fig.3 within
 the experimental resolution.
\end{references}
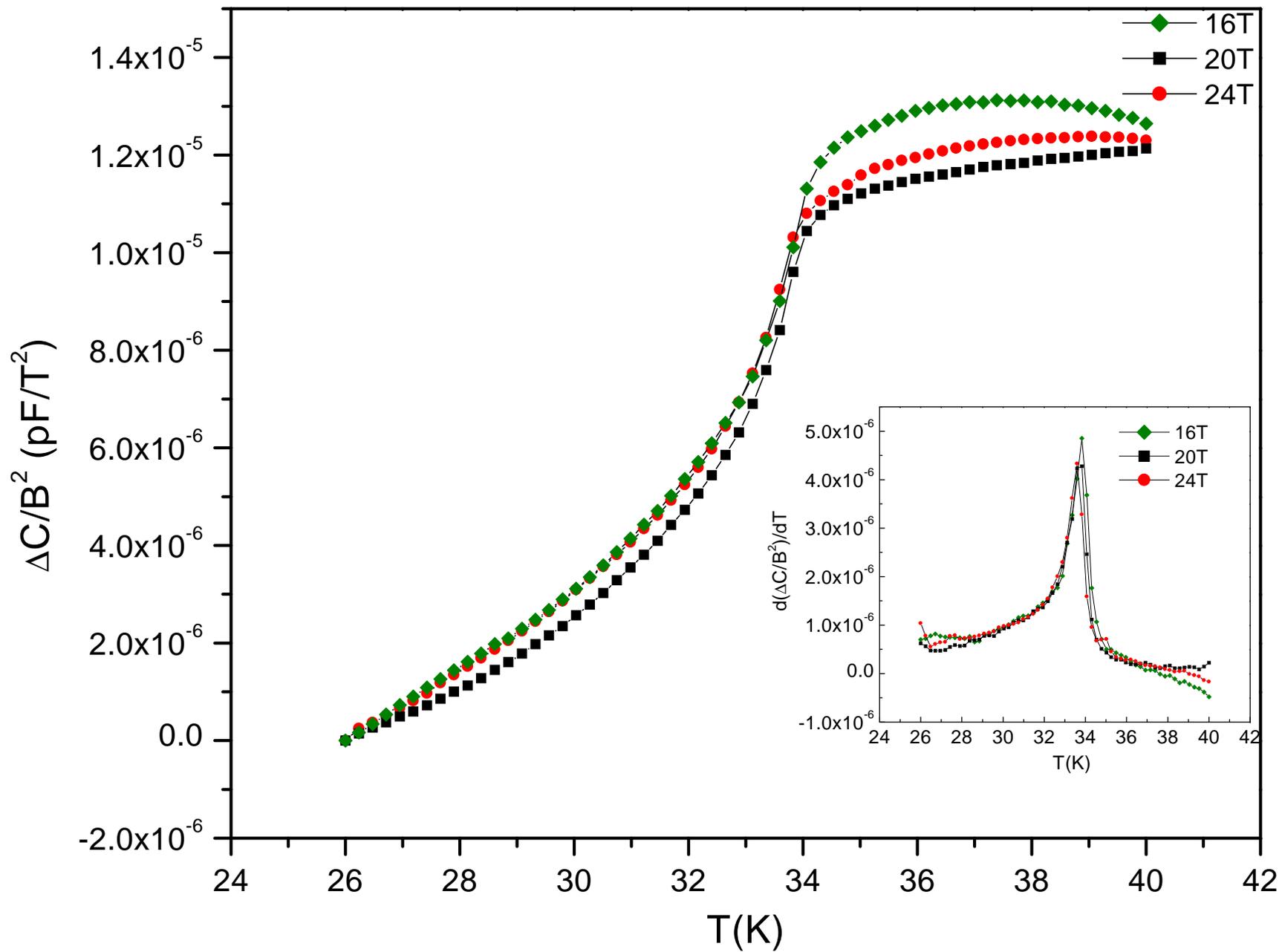
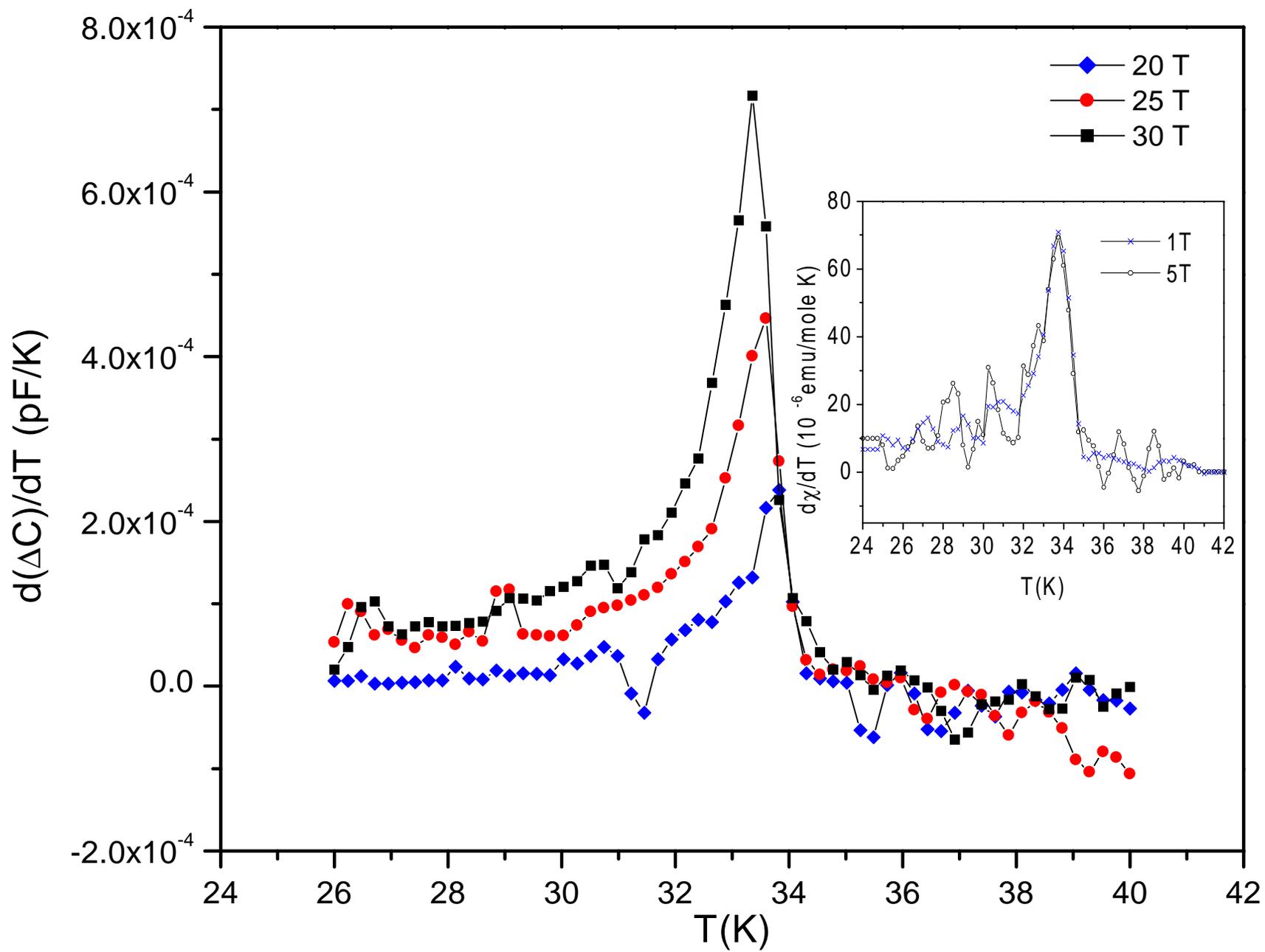
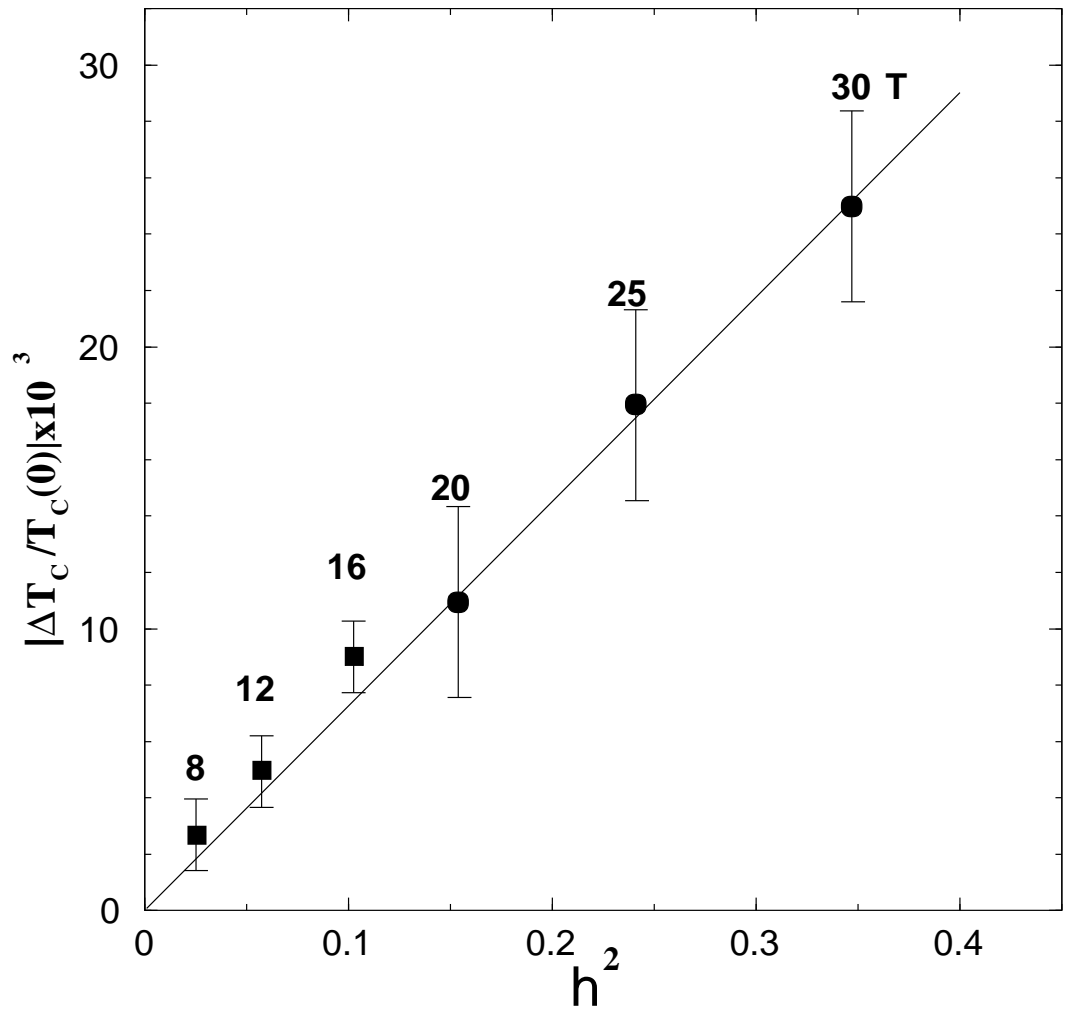
\end{document}